\documentclass[final,1p,times]{elsarticle}

\usepackage{amssymb}
\usepackage{graphicx}

\usepackage{lineno}
\newcommand{\geant}       {\textsc{Geant4}}
\newcommand{\dance}       {Dance \emph{et al.}}
\newcommand{\ambrose}     {Ambrose \emph{et al.}}

\journal{Nuclear Instruments and Methods B}

\begin{document}

\begin{frontmatter}



\title{Validation of the Geant4 simulation of bremsstrahlung from thick targets below 3~MeV}


\author[lns,lngs]{L.~Pandola}\ead{pandola@lns.infn.it}
\address[lns]{INFN, Laboratori Nazionali del Sud, Via Santa Sofia 62, I-95125 Catania, Italy}
\address[lngs]{INFN, Laboratori Nazionali del Gran Sasso, S.S. 17 bis km 18+910, I-67100 Assergi (AQ), Italy}
\author[inail]{C.~Andenna} 
\address[inail]{INAIL, Dipartimento Innovazioni Tecnologiche e 
Sicurezza degli Impianti, Prodotti ed Insediamenti Antropici, Via Alessandria 
220, I-00198 Roma, Italy}
\author[iss]{B.~Caccia}
\address[iss]{Dipartimento Tecnologie e Salute, Istituto Superiore di Sanit\`a 
and INFN, Gruppo Collegato dell'Istituto Superiore di Sanit\`a, 
Viale Regina Elena 299, I-00161 Roma, Italy}


\begin{abstract}
The bremsstrahlung spectra produced by electrons impinging on thick targets 
are simulated using the \geant\ Monte Carlo toolkit. Simulations are 
validated against experimental data available in literature for a range of 
energy between 0.5 and 2.8~MeV for Al and Fe targets and for a value of energy of 
70~keV for Al, Ag, W and Pb targets. All three independent sets of electromagnetic 
models available in \geant\ to simulate bremsstrahlung are tested. 
A quantitative analysis is performed reproducing with each model the energy 
spectrum for the different configurations of emission angles, energies and targets. \\
At higher energies (0.5-2.8~MeV) of the impinging electrons on Al and Fe targets, 
\geant\ is able to reproduce the spectral shapes and the integral photon emission 
in the forward direction. The agreement is within 10-30\%, depending on 
energy, emission angle and target material. The physics model based on the Penelope 
Monte Carlo code is in slightly better agreement with the measured data than the other two. 
However, all models over-estimate the photon emission in the backward hemisphere.
For the lower energy study (70~keV), which includes higher-Z targets, all models  
systematically under-estimate the total photon yield, while still providing a 
reasonable agreement between 10 and 50\%.\\
The results of this work are of potential interest for medical physics applications, 
where knowledge of the energy spectra and angular distributions of photons is needed 
for accurate dose calculations with Monte Carlo and other fluence-based methods. 
\end{abstract} 


\end{frontmatter}


\section{Introduction}
The need for the precise description of the photon emission by electron 
bremsstrahlung in thick targets is common to many fields of research, including 
medical physics, astrophysics and astroparticle physics. While bremsstrahlung 
starts to dominate over ionization for energies of tens of MeV, the 
process may be relevant and measurable also at much lower energies, because of the 
longer mean free path of photons with respect to electrons.\\
The simulation of the bremsstrahlung emission is especially  
relevant in medical physics applications, 
where the clinical x-ray beams are produced by electrons of kinetic energies 
between 10~keV and 50~MeV decelerated in metallic targets. A fraction of the electron’s 
kinetic energy is transformed in the target into heat, and a fraction of the energy is  
emitted in the form of bremsstrahlung photons. The knowledge of the energy spectrum and angular 
distribution of photon beams produced in the interaction is essential for accurate dose 
calculations in the patient and represents the most rigorous description of beam 
quality~\cite{faddegon1990}. In external radiotherapy, to optimize the dose distribution in 
a patient, the beam is modulated in intensity and shape. 
The radiation treatment outcome is related to the accuracy in the delivered dose to 
the patient that depends on the accuracy of beam data~\cite{TG-106} and quality assurance 
procedures~\cite{AAPM40,TG-142}. The quality assurance program for linear accelerators requires 
that the machine characteristics do not deviate significantly from their baseline
values acquired at the time of acceptance and commissioning~\cite{AAPM40,TG-142}.
The main reason for the requirement of high accuracy in dose delivery is typically the narrow 
margin between the dose needed for tumor control and the dose causing complications for healthy 
tissues. Targets in clinical linear accelerators are thick enough to stop the primary 
electrons completely.\\ 
Due to the interplay with other concurrent physical processes 
which can affect the kinetic energy and the direction of the electrons in thick media,  
it is very difficult to obtain the spectral and angular photon distributions 
using analytical methods such as the Schiff theory~\cite{dist2bs,Desobry}.
Calculations must hence be performed by using Monte Carlo simulations. Many different 
codes are available for this purpose, including 
\geant~\cite{geant1,geant2}, Penelope~\cite{pen1}, EGSNRC~\cite{egs} 
and Fluka~\cite{fluka,fluka2}. \\
The aim of this work is the validation of the bremsstrahlung emission (total radiated energy and 
energy/angular spectra) in thick 
targets as predicted by \geant, tailored to the requirements 
and the applications that are typical of medical physics. Since the region between 
15 and 30 MeV was already considered by other authors~\cite{faddegon}, a 
special focus is given to lower energies. Suitable experimental 
measurements of bremsstrahlung emission in thick targets at such low energies are not 
readily found in the literature. Two sets of measurements are considered for this work: 
\begin{enumerate}
\item \dance~\cite{dance}, which reports absolute energy and angular double-differential 
photon spectra for electron energies between 0.5 and 2.8~MeV in aluminum and iron thick targets; 
\item \ambrose~\cite{ambrose}, which displays absolute energy spectra for 70 keV electrons 
impinging at two different angles on thick targets of aluminum, silver, tungsten and 
lead\footnote{Measurements in others materials were also taken (``for a range of Z from 6 to 92''). 
However, they are not reported or shown individually in the paper and are hence not usable 
for validation purposes.}. 
\end{enumerate}
The work presented here is not meant to be a comprehensive or a quantitative 
validation of the Monte Carlo simulation,
as done in other papers in the recent literature~\cite{pia1,pia2,pia3,pia4}. 
The main goal is to give a 
general overview of the physics performance of \geant\ in this domain as well as    
to clearly indicate the regimes where measurement and simulation disagree, 
which may be relevant to the users of the code.\\

The paper is organized as follows: in Sect.~\ref{sec:mcgen} the \geant-based 
Monte Carlo simulation which was developed to reproduce the reference experimental data is described. 
An overview is given of the alternative physics models that are presently available in \geant\ to 
describe bremsstrahlung in the energy range of interest. The outcome and the results provided 
by the simulation are shown and reported in Sect.~\ref{sec:disc}; the agreement with the 
reference data is discussed in detail and strong and weak points of the simulations are 
emphasized. General conclusions about the validity and reliability of \geant-based simulations 
concerning the production of bremsstrahlung photons are drawn in Sect.~\ref{sec:conclusions}, 
with a focus on medical applications.

\section{The Monte Carlo simulation with Geant4} \label{sec:mcgen}
As mentioned above, data from Refs.~\cite{dance,ambrose} were used in this work 
to validate the physics results provided by the \geant\ code. 
\textsc{Geant4}~\cite{geant1,geant2} is a general-purpose toolkit for the 
Monte Carlo simulation of the propagation of particles in matter. It includes 
a variety of physics models to describe the electromagnetic and hadronic 
interactions of many kinds of particles, including $\gamma$-rays, leptons, 
baryons, mesons and nuclei. The comparison work presented in this paper 
has been carried out with the version 9.6.p02 of \geant\ (May, 2013). While a 
more recent version of \geant\ was made available in the meanwhile (10.0, 
December, 2013), no changes are reported in the release notes for the physics 
models that are relevant for the simulation of bremsstrahlung photons at low 
energy.

\subsection{Geant4 physics models} \label{sec:physics}
Three independent sets of electromagnetic models are available 
in \geant\ that are appropriate to simulate bremsstrahlung 
in the energy range considered in this work: 
``Standard''~\cite{geant1,std,opt3,stdlowe,stdlowe2}, 
``Livermore''~\cite{stdlowe,stdlowe2,livermore} and 
Penelope~\cite{pen1,opt3,stdlowe2}. The constructors available in \geant\ 
can be used to register all electromagnetic processes in the 
user application. In the case of the Standard package the so-called 
``Option3'' (\texttt{G4EmStandardPhysics\_option3}) is used which is tailored 
to medical 
and space application~\cite{opt3}. For Livermore and Penelope 
the constructors \texttt{G4EmLivermorePhysics} and 
\texttt{G4EmPenelopePhysics} are used, 
respectively. All model parameters are unchanged with respect to the 
default provided in \geant\ 9.6.p02.\\
In the \geant\ unified scheme for the electromagnetic physics, 
different models can be used for the same physics process. The process 
which is devoted to describe the bremsstrahlung of electrons and positrons 
is \texttt{G4eBremsstrahlung}. \\
The bremsstrahlung models used by \texttt{G4eBremsstrahlung} for the physics 
lists used in this work are: \texttt{G4SeltzerBergerModel} in 
Option3\footnote{For electrons and positrons above 1~GeV, a 
specific relativistic model \texttt{G4eBremsstrahlungRelModel} is used. 
Given the energy range of interest, this is not relevant for the present 
work, and actually never invoked.}, 
\texttt{G4LivermoreBremsstrahlungModel} in ``Livermore'' 
and \texttt{G4PenelopeBremsstrahlungModel} in ``Penelope''.\\
The cross sections used by the model \texttt{G4SeltzerBergerModel} are based 
on the interpolation of published tables~\cite{sel1,sel2}, which account 
for the bremsstrahlung emission in the field of nuclei and of atomic 
electrons. The published tables contain energy-differential cross sections 
$d\sigma/dE$ between 1~keV and 10~GeV.
The uncertainty reported in the original work for the energy range of 
interest of this work is ``about 10\%'' below 2~MeV and ``between 5\% and 
10\%'' between 2~MeV and 50~MeV~\cite{sel2}.
The integral cross section is calculated numerically from the 
differential tables at the initialization of \geant. The 
angular distribution, following the successful sampling of the radiated 
energy, is calculated according to a simplified version of the Tsai's 
formula~\cite{tsai1,tsai2}, as described in Ref.~\cite{refman}. The simplified 
formula is appropriate for very high energies, but is expected to be much 
less accurate in the MeV range.\\
The total cross section used by the model 
\texttt{G4LivermoreBremsstrahlungModel} is obtained from the 
interpolation of the evaluated cross section data from the EEDL Livermore 
library~\cite{liv}.  The shape of the photon energy 
spectra is also derived by interpolation of the data reported from the 
EEDL library. Values are listed as a function of the ratio $\kappa$ between 
the photon energy $k$ and the electron energy $T_0$: the tabulated data set 
contains 15 points for each element ranging between $\kappa=0.01$ and $\kappa=1$, 
and a linear interpolation method is used.
The angular distribution is sampled according to 2BS 
formula by Koch and Motz~\cite{dist2bs}, using the algorithm developed by 
Bielajew~\cite{bie} and implemented in EGS4. The 2BN formula by 
Koch and Motz~\cite{dist2bs} and a simplified version of the 
Tsai distribution~\cite{tsai1} are also available as alternative options.
In the present work, only the 2BS default generator was considered. \\
The model \texttt{G4PenelopeBremsstrahlungModel} is the re-engineering in 
\geant\ of the bremsstrahlung model of  
Penelope Monte Carlo code v2008~\cite{pen1,pen2,pen3}. The total cross sections that are 
used for the evaluation of the restricted cross sections above the production 
threshold are taken from the EEDL Livermore library. The shape of 
the photon emission spectra are sampled according to a parametrization of the 
data reported in the tables of Ref.~\cite{sel2}. Data are available for all 
elements; the grid has 57 electron energies between 1~keV and 10~GeV and 
32 points in the photon energy $\kappa = k/T_0$ between $10^{-12}$ and 1. A 
log-log interpolation is performed between the grid points. The angular 
distribution 
is sampled from a modified Lorentz distribution~\cite{pen1,pen3}, whose 
parameters are fitted to match the shape functions reported by Kissel~\cite{kissel} 
for the following benchmark cases: $Z$=2, 8, 13, 47, 79 and 92; $E$=1, 5, 
10, 50, 100 and 500 keV; $\kappa$=0, 0.6, 0.8 and 1.0.  \\
The abstract interface \texttt{G4VEmAngularGenerator} provides the possibility 
of using for each model an alternative angular generator than the default 
one. For instance, the angular generator \texttt{G4PenelopeBremsstrahlungAngular} 
which implements the sampling according to the modified Lorentz dipole distribution 
can be also registered to the other models.

\subsection{The user application}
A dedicated user application was developed to simulate the passage of electrons 
in a thick target, and to score energy and direction of all photons leaving 
the target. All relevant physics processes were considered by using the 
\geant\ constructors described in Sect.~\ref{sec:physics}. Therefore, the
simulation properly accounts for the slow-down of the electrons due to 
ionization and for the possible back-scattering from the target. 
The energy $T_0$ of the impinging electrons, the thickness and 
the material of the target can be changed at run-time by means of a user 
interface. In the simulation jobs for \dance\ the  
cut-in-range for the production of secondary photons and electrons was set to 2.5~$\mu$m, 
which is much smaller than the thickness of the targets (between 0.32 and 6.4~mm in 
the various configurations). The minimum photon energy which is allowed by the cuts 
was 250~eV in both Al and Fe, which is below the  
energy of the characteristic x-rays. The cut values are also 
well below the experimental energy 
threshold reported in \dance~\cite{dance}, which is between 46 and 171~keV. Such a cut 
was found to be the optimal trade-off between CPU performances and tracking precision 
in the target. In the simulation jobs for \ambrose\ the production cut-in-range is 
1~$\mu$m, while the thickness of the target is between 18 and 94 $\mu$m. Also in this case,
the threshold used in the simulation was low enough to allow the production of fluorescence 
x-rays in all target materials.\\
Simulations were run on different Linux machines, in which \geant\ had been built 
and compiled from the source code. 
At least $10^9$ primary electrons were generated for each of the eight configurations 
reported by \dance, i.e. four energies ($T_0$ = 0.5, 1.0, 2.0 and 2.8~MeV) and two 
target materials (Al and Fe), and for the four configurations reported by \ambrose\  
(i.e. Al, Ag, W and Pb). Primary electrons were assumed to form an ideally mono-chromatic 
pencil beam. 
Simulations were then run for the three sets of \geant\ physics 
models described in Sect.~\ref{sec:physics}. The CPU time required to get comparable 
statistical precision in the simulation outputs was typically between 8 and 24 hours
for each specific configuration. Time differences are mostly related to the target 
thickness, than to the physical models considered: for a given configuration, the 
Penelope and Livermore simulations take 15-20\% longer than Option3. \\

\section{Comparison with experimental data: results and discussion} \label{sec:disc}
The papers that were selected as the reference for this study 
do not provide tabulated data: measurements are displayed in graphical form only. 
No assessment of experimental uncertainties is available from \ambrose\ The 
uncertainty of each data point in the double differential spectra by \dance\ is evaluated 
by the authors to be between 15\% and 18\%. The uncertainty due to the digitization procedure 
is much smaller ($< 5\%$), so that the global uncertainty considered here for the digitized data 
points is 20\%. The uncertainty of the integrated radiated intensity reported in \dance, as 
inferred from the size of the error bars in the original plots, is about 11\%. \\

\subsection{Electrons between 0.5 and 2.0 MeV on aluminum and iron targets} \label{sect:dance}
\begin{figure*}[tbp] 
\includegraphics[width=0.95\columnwidth]{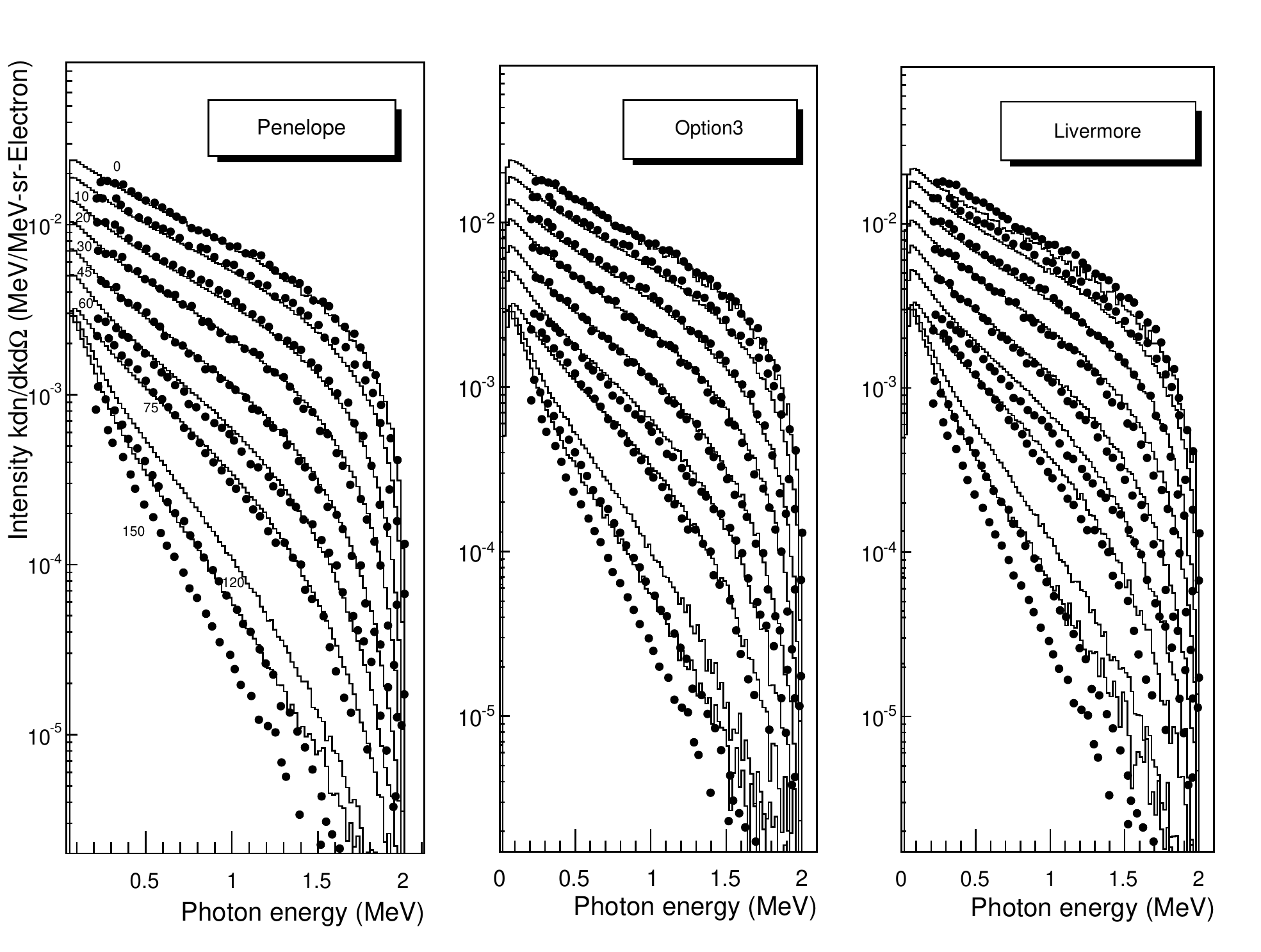} 
\caption{\label{fig::allAl} Absolute energy spectra of photons emitted by 2-MeV electrons 
impinging on an aluminum target (1.20~g/cm$^2$) at nine different angles (0, 10, 20, 30, 45, 60, 
75, 120 and 150 degrees). Black circles are experimental data from \dance\ and solid histogram 
are the corresponding Monte Carlo simulations. The three sets of models available in \geant\ 
are considered. }
\end{figure*}

Fig.~\ref{fig::allAl} displays the absolute double-differential distributions (in energy and 
angle) for the emitted photons in one of the configurations reported by 
\dance, 2.0~MeV electrons 
on aluminum. 
Experimental data are superimposed to the results of the \geant\ simulations with the 
three sets of physics models at nine angles between 0 and 150 deg.  
All three models predict the correct absolute scale and are able to reproduce 
the shape of the energy spectra at forward emission angles, which make 
the leading contribution to the total radiated energy. Nevertheless, all models over-estimate the 
bremsstrahlung emission in the backward hemisphere and predict a harder energy distribution 
than measured. The Penelope and Option3 models provide very comparable results and 
in good agreement with data up to $\sim 75$~deg. The Livermore model provides a very good 
agreement at intermediate angles, while cannot reproduce data at very small and very large angles. 
The features described above are observed, at smaller or larger extent, in all configurations 
of \dance, between 0.5 and 2.8~MeV. \\
\begin{figure}[tbp] 
\includegraphics[width=0.95\columnwidth]{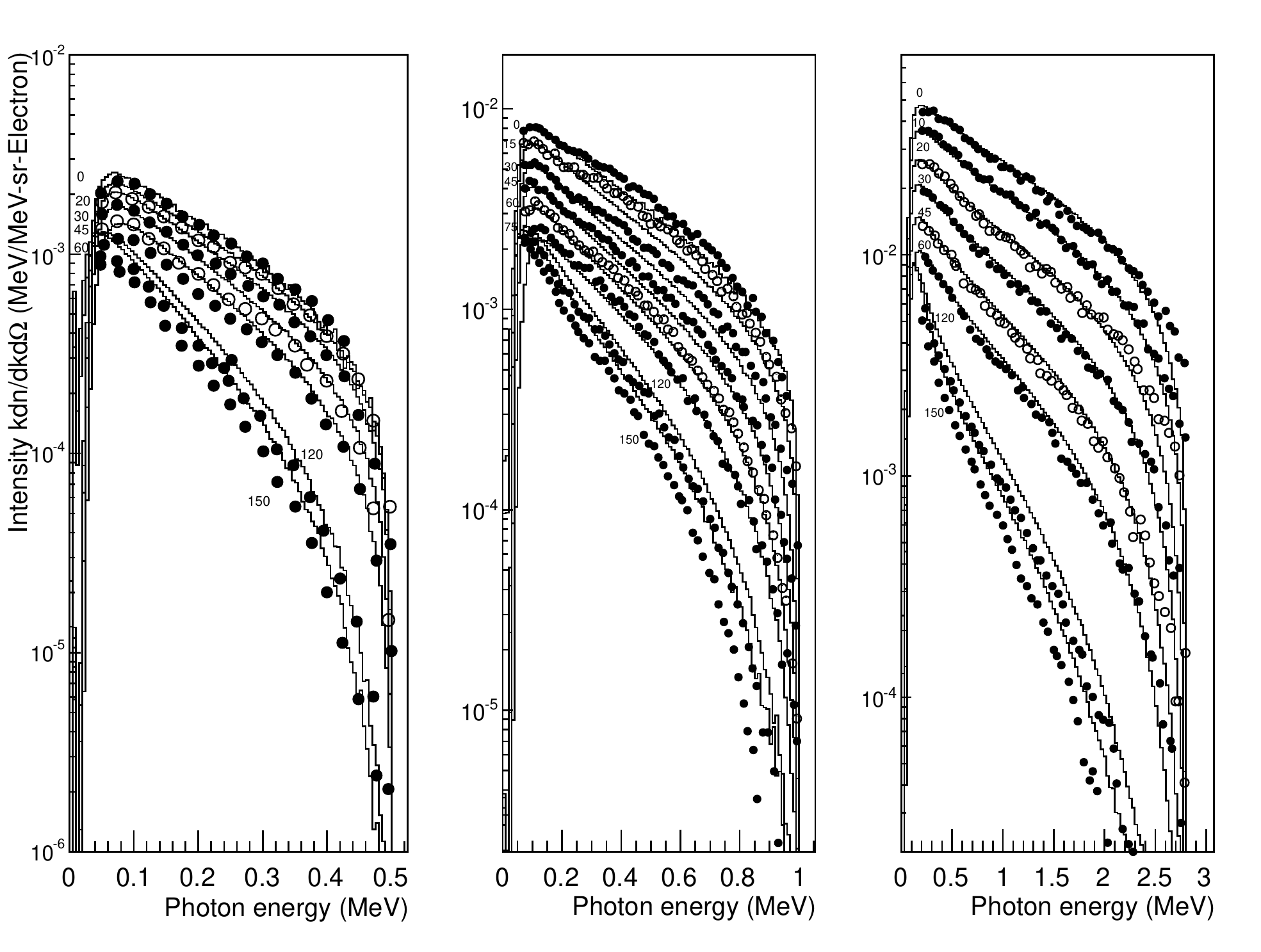} 
\caption{\label{fig:penelopeFe} Absolute energy spectra of photons emitted by electrons of 
0.5~MeV (left), 1.0~MeV (middle) and 2.8~MeV (right) on iron targets, at different angles 
between 0 and 150 deg. Circles are experimental data from \dance\ and solid histograms 
are the \geant\ simulations obtained with the Penelope model. Some sets of data points 
are shown as empty circles for visual clarity. The thickness of the targets is 
0.257~g/cm$^2$ (0.5~MeV beam), 0.613~g/cm$^2$ (1.0~MeV beam) and 2.31~g/cm$^2$ 
(2.8~MeV beam).}
\end{figure}
The dependence of the double-differential distributions on the electron energy $T_0$ is 
displayed in Fig.~\ref{fig:penelopeFe} for the Penelope model: configurations 
with Fe targets from \dance\ are considered, with energy between 0.5 and 2.8~MeV. While 
the general comments above still hold (i.e. correct absolute scale, correct shape of 
energy spectra at forward angles) it is apparent that the agreement with data gets 
worse at lower energies. Such a feature is common also to the other physics 
models. Similarly, all models give results closer to the reference data for 
lower Z: the agreement with measurements is systematically better in the Al target 
configurations than in the Fe target.\\
\begin{figure}[tbp] 
\includegraphics[width=0.95\columnwidth]{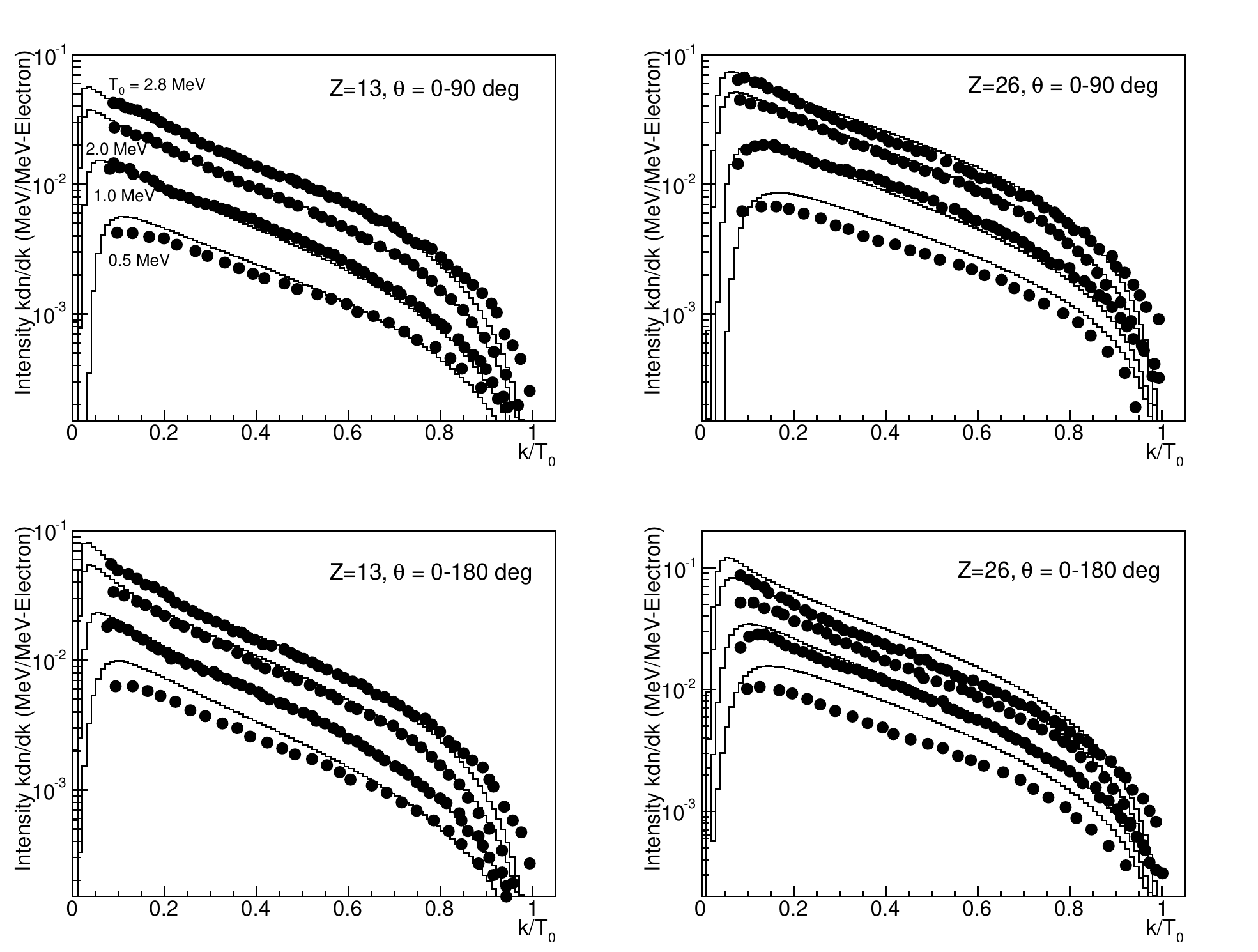} 
\caption{\label{fig:penelope1D} Absolute energy spectra of photons emitted by electrons 
of kinetic energy $T_0 = 0.5, 1.0, 2.0$ and 2.8~MeV on aluminum (left column) and iron 
(right column) targets. The upper row shows the spectra restricted to forward emission 
only, and the lower row to the full space. Circles are experimental data from \dance\ 
and solid histograms are the \geant\ simulations obtained with the Penelope model. Photon 
energy on the x-axis is normalized with respect to the kinetic energy of the primary 
electrons.}
\end{figure}
The same trend is also visible in Fig.~\ref{fig:penelope1D}, which displays the single-differential 
photon energy spectra obtained with the Penelope model at all electron energies $T_0$ 
and in both targets by \dance : Al ($Z=13$) and Fe ($Z=26$). 
Spectra are shown separately for the forward emission hemisphere only (upper row) 
and for the full space (lower row). A very good agreement is observed for the photons emitted by  
higher energy electrons and in the forward hemisphere. The agreement is spoiled for lower electron energy 
and higher $Z$ targets. As discussed above, an over-production of bremsstrahlung photons is observed in 
the backward hemisphere, which is more evident at lower energy and higher Z. \\
The paper by \dance\ also reports (in graphical form) the photon radiated energy vs. the bombarding 
energy $T_0$, integrated either over all angles or over the forward hemisphere only.
Table~\ref{tab:yields} summarizes the experimental data $I_{meas}$ and the ratio $I_{sim}/I_{meas}$ 
between Monte Carlo predictions and data. The uncertainty $\sigma_{meas}$ of the data 
points $I_{meas}$ from \dance\ is about 11\% in all configurations.
Results for forward emission are very satisfactory: all models agree to 
data for all eight configurations of Tab.~\ref{tab:yields}. The Livermore 
and Penelope predictions are within 10\% in all cases, except for Fe at 0.5 MeV (lowest 
energy at highest Z); the two models are practically equivalent for the aluminum target, while 
Penelope provides a slightly better agreement with the iron target data. Option3 agrees with 
data within 30\%. Table~\ref{tab:yields} also reports the $X^2$ calculated as
\begin{equation}
X^2 = \sum \frac{(I_{meas} - I_{sim})^2}{\sigma^2_{meas}}
\end{equation} 
in order to quantify the global performance of the three sets of models\footnote{The $X^2$ reported here 
is calculated in the assumption that uncertainties $\sigma_{meas}$ of the data points are statistical 
and uncorrelated. 
No information is reported in \dance\ about the correlation of experimental errors and about the 
relative contribution of systematic uncertainties. Therefore, the $X^2$ should not be regarded 
here as a $\chi^2$ in its strict statistical meaning, but rather be seen as a parameter to assess the relative 
performance of the models.}. The Penelope and the Livermore models 
are both consistent with the data, with the former providing a slightly better agreement.
The Option3 model, while giving a very good agreement for the shape of the double-differential spectra, 
seems to be the least accurate in the prediction on the total energy irradiated in the forward 
direction.\\ 
As expected, results are less good if photons irradiated in the backward direction 
are also considered. All models fail to reproduce the data point of Fe at 0.5~MeV at more than 
3$\sigma_{meas}$ and are inconsistent (on average) with the reference data. The single 
data point of Fe at 0.5 MeV drives the high $X^2$ values; if this problematic measurement is 
excluded, all models agree with data within 15-20\%. Nevertheless, they show a systematic 
over-estimate of the total radiated energy. The three sets are practically equivalent in the 
prediction of the all-space yield, with Penelope still providing a bit better agreement with 
measurements than the others and Option3 being the next.\\

%
\begin{table*}[htbp]
\centering \caption{\label{tab:yields} Total energy radiated in photons, normalized per primary electron, 
for all the energy-target configurations reported in the paper by \dance. The uncertainty of experimental 
data points is approximately 11\% for all configurations.
The last three columns summarize the ratio between the photon intensity predicted by the \geant\ simulation 
(Option3, Livermore and Penelope models, respectively) and experimental data. The upper half of the 
table reports the photon yield restricted only to the forward 
hemisphere, while the lower half reports the all-space yield. The agreement between experimental 
data and simulations is assessed by a $X^2$ statistics. The number $\nu$ of degrees of freedom 
is also reported.}
\begin{tabular}{cc|c|lll}
Material & Energy & Data & \multicolumn{3}{c}{Simulation/data} \\
 & (MeV) & (MeV/electron) & Option3 & Livermore & Penelope \\
 &       &  ($\pm 11\%$) & \multicolumn{3}{c}{} \\
\hline
\multicolumn{2}{l}{forward ($\theta < \pi/2$)} \\
\hline
Al & 0.5 & $8.80 \cdot 10^{-4}$ & 0.99 & 1.00 & 1.01 \\
Al & 1.0 & $4.45 \cdot 10^{-3}$ & 0.70 & 0.90 & 0.93 \\
Al & 2.0 & $1.65 \cdot 10^{-2}$ & 1.00 & 1.02 & 0.99 \\
Al & 2.8 & $3.52 \cdot 10^{-2}$ & 0.98 & 1.00 & 0.97 \\
Fe & 0.5 & $1.41 \cdot 10^{-3}$ & 1.27 & 1.24 & 1.23 \\
Fe & 1.0 & $7.94 \cdot 10^{-3}$ & 0.83 & 0.93 & 0.91 \\
Fe & 2.0 & $2.99 \cdot 10^{-2}$ & 0.90 & 1.04 & 1.01 \\
Fe & 2.8 & $6.05 \cdot 10^{-2}$ & 0.99 & 1.03 & 1.00 \\
\multicolumn{2}{l|}{$X^2$ ($\nu=8$)} & & 18.2 & 9.8 & 6.3 \\
\hline
\multicolumn{2}{l}{all space ($\theta < \pi$)} \\ 
\hline
Al & 0.5 & $1.15 \cdot 10^{-3}$ & 1.16 & 1.18 & 1.16 \\
Al & 1.0 & $5.20 \cdot 10^{-3}$ & 0.81 & 1.06 & 1.08 \\
Al & 2.0 & $1.78 \cdot 10^{-3}$ & 1.11 & 1.15 & 1.11 \\
Al & 2.8 & $3.98 \cdot 10^{-2}$ & 0.99 & 1.03 & 0.99 \\
Fe & 0.5 & $2.08 \cdot 10^{-3}$ & 1.34 & 1.37 & 1.35 \\
Fe & 1.0 & $1.03 \cdot 10^{-2}$ & 0.94 & 1.10 & 1.08 \\
Fe & 2.0 & $3.65 \cdot 10^{-2}$ & 0.99 & 1.15 & 1.13 \\
Fe & 2.8 & $7.52 \cdot 10^{-2}$ & 1.05 & 1.09 & 1.05 \\
\multicolumn{2}{l|}{$X^2$ ($\nu=8$)} & & 21.9 & 26.7 & 19.7 \\
\multicolumn{2}{l|}{$X^2$ w/o Fe 0.5 MeV ($\nu=7$)} & & 7.9 & 9.8 & 5.4 \\
\hline
\end{tabular}
\end{table*}
%
%

\subsection{70 keV electrons on aluminum, silver, tungsten and lead targets}
\begin{figure}[tbp] 
\includegraphics[width=0.95\columnwidth]{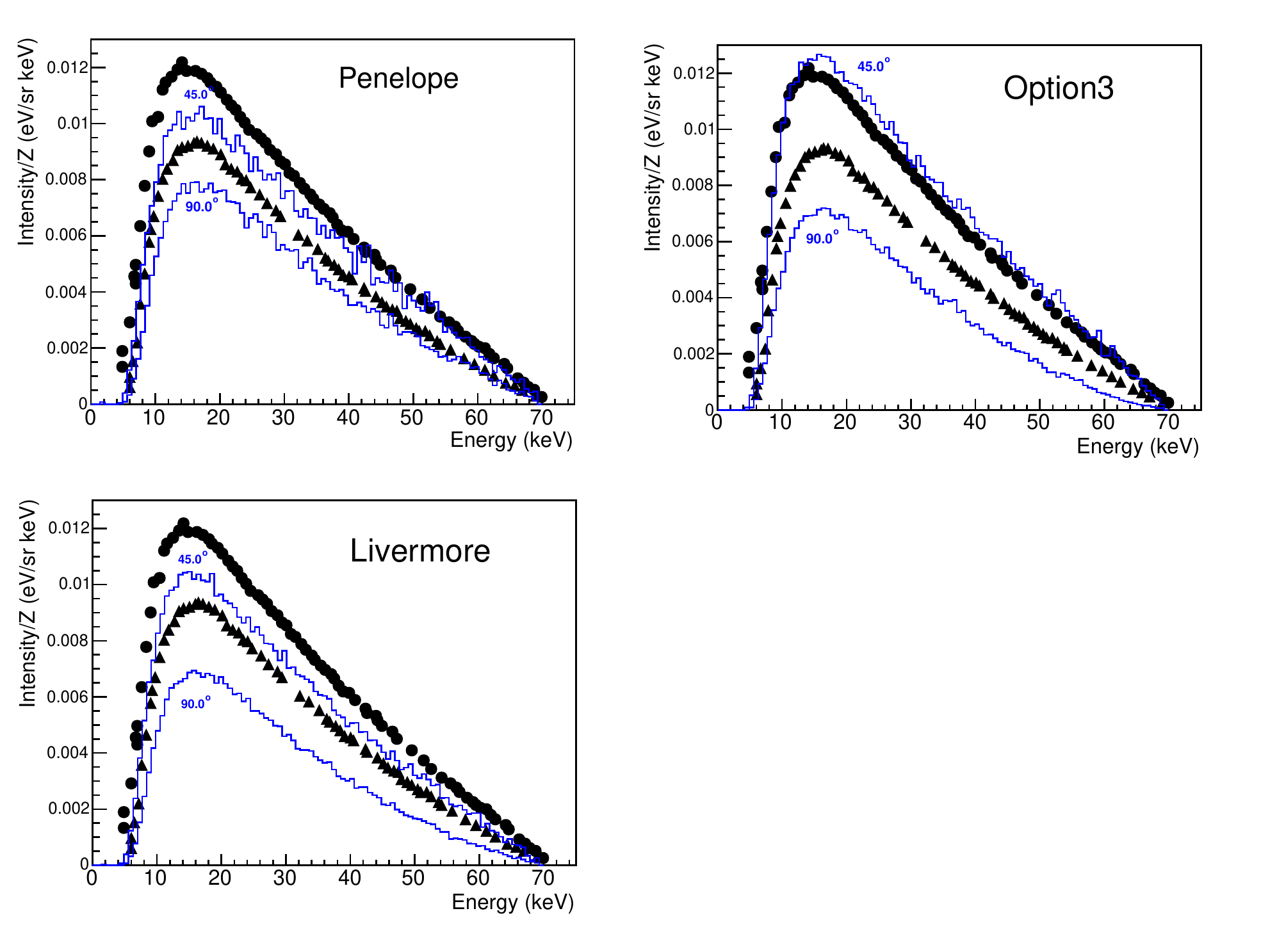} 
\caption{\label{fig:ambroseModels} Absolute energy spectra 
$\frac{k}{T_0 Z} \frac{dN}{dk d\Omega}$ of photons emitted by 70~keV 
electrons impinging on a 25.4~mg/cm$^2$ aluminum target at 45 and 90 degrees. Spectra 
are normalized according to atomic number $Z$, solid angle and incident energy $T_0$.
Circles and triangles are experimental data from \ambrose\ (45 and 90 degrees, respectively). 
Solid histograms are the \geant\ simulations obtained with Penelope (upper left), Option3 
(upper right) and Livermore (lower left) models.}  
\end{figure}

\begin{figure}[tbp] 
\includegraphics[width=0.95\columnwidth]{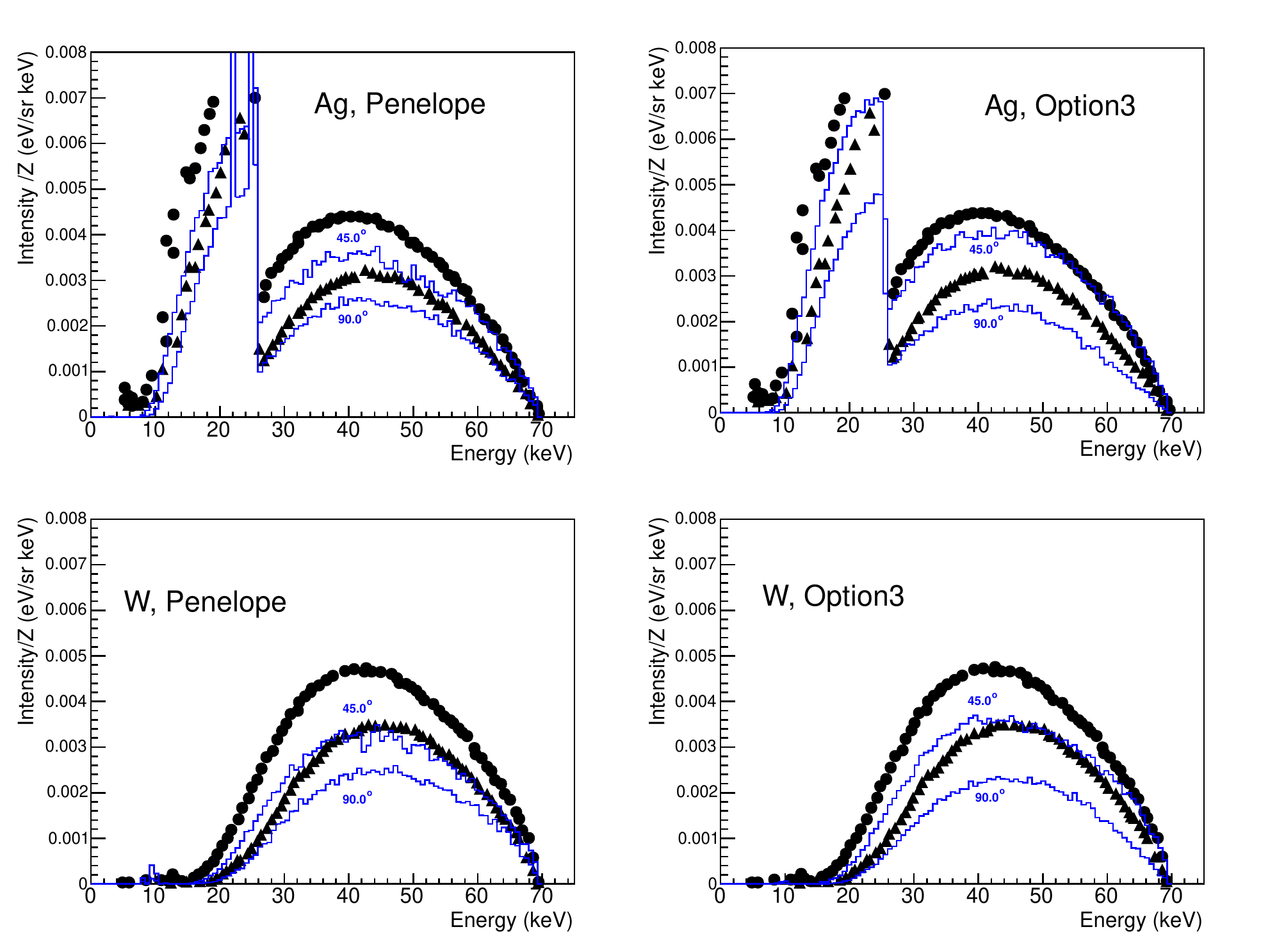} 
\caption{\label{fig:ambroseMaterials} Absolute energy spectra $\frac{k}{T_0 Z} \frac{dN}{dk d\Omega}$ 
of photons emitted by 
70~keV electrons impinging on silver (28.06~mg/cm$^2$) and tungsten (50.7~mg/cm$^2$) 
targets at 45 and 90 degrees. 
Spectra are normalized according to atomic number $Z$, solid angle and 
incident energy $T_0$. Circles and triangles are experimental data from \ambrose\ 
(45 and 90 degrees, respectively). Solid histograms are the \geant\ simulations 
obtained at the two angles with Penelope (left column) and Option3 (right column) 
models. }  
\end{figure}

The paper from \ambrose\ reports measurements of photon spectra emitted 
by 70 keV electrons impinging on thick targets of various materials. 
The targets are rotated at 45 deg with respect to the incident beam and they are 
thick just enough to stop the incident electrons. Photons are detected at emission 
angles of 45 and 90 degrees with respect to the beam.
At the low energy considered by \ambrose\ the electron ionization process largely 
dominates and the production of bremsstrahlung photons is relatively 
scarce\footnote{Still, fluorescence x-rays are 
produced following electron ionization and the characteristic peaks are well visible 
in the spectra.}.
Nevertheless, bremsstrahlung in this energy range is of potential interest for 
specific medical applications, like imaging and superficial radiotherapy (e.g. 
for skin lesions).\\
The validation of \geant\ with the data at 70~keV has the advantage that bremsstrahlung 
models are tested in an ``unusual'' energy range, where the process is rare, and 
experimental measurements are scarce. In fact, bremsstrahlung models are typically 
developed and tailored for energies above 1~MeV: the capability to reproduce 
measurements taken at a much different energy cannot be taken for granted a priori 
and thus provides a confirmation of the validity of the underlying physics modeling.
Figure~\ref{fig:ambroseModels} shows the absolute comparison between experimental data 
and \geant\ simulations for both emission angles in the aluminum target, as obtained 
with the three sets of physics models. All models 
slightly under-estimate the experimental values, with the exception of the curve obtained 
at 45 deg with Option3. Similar conclusions hold for higher-Z targets, as displayed in 
Fig.~\ref{fig:ambroseMaterials} (Penelope and Option3 only). 
\begin{figure}[tbp] 
\includegraphics[width=0.95\columnwidth]{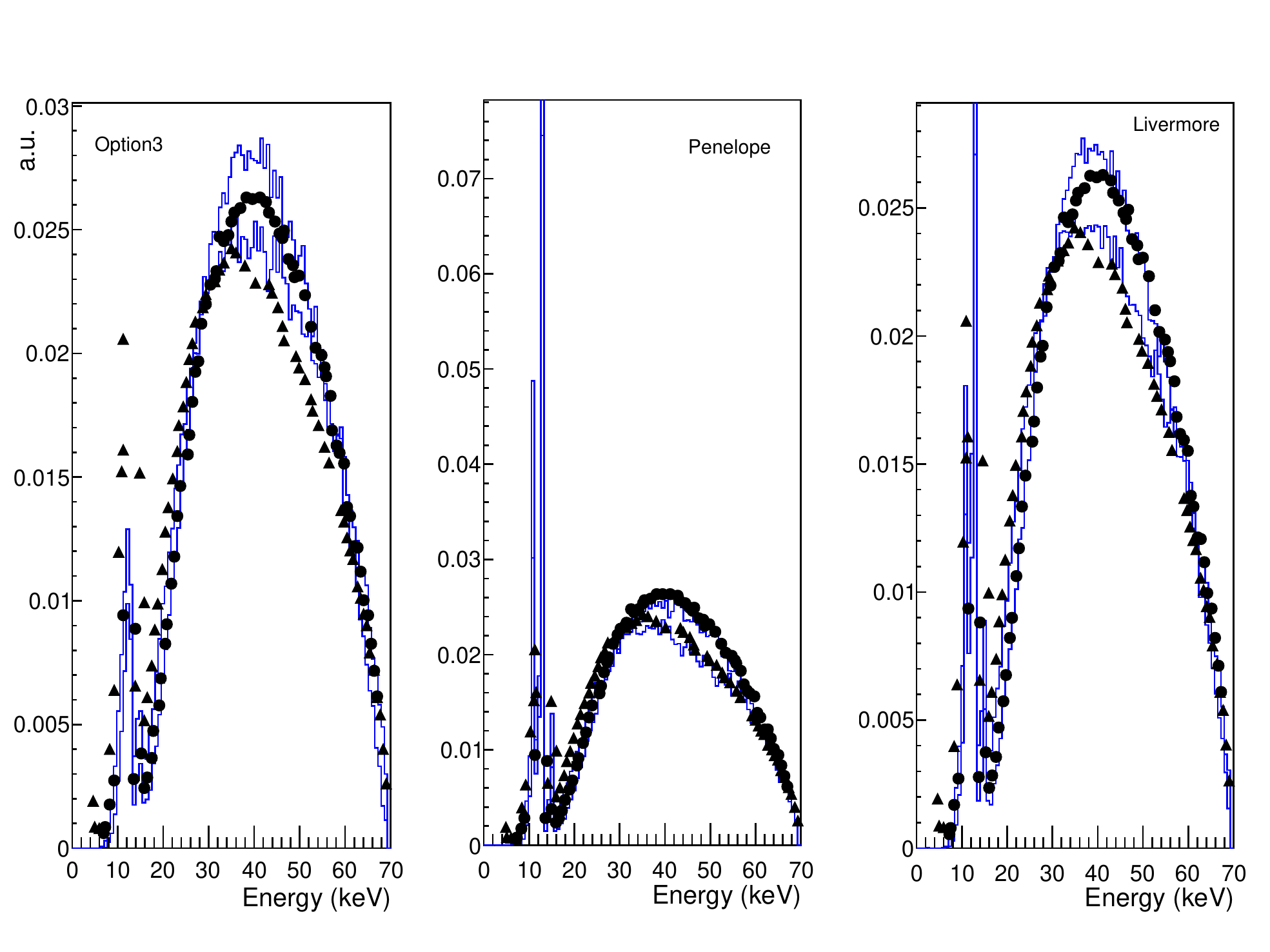} 
\caption{\label{fig:ambroseNormalized} Qualitative comparison of normalized energy spectra 
of photons emitted by 70-keV 
electrons impinging on a 20.89~mg/cm$^2$ Pb target at 45 and 90 degrees.
Circles and triangles are experimental data from \ambrose\ (45 and 90 degrees, respectively) 
Solid histograms are the \geant\ simulations obtained with Option3 (left), Penelope (middle) 
and Livermore (right) models. Histograms are normalized to match the integrated intensity 
of the bremsstrahlung continuum ($> 20$~keV) from the experimental data.}  
\end{figure}
Nevertheless, the absolute agreement between data and models concerning the integral 
photon yield is relatively good: models are able to reproduce the measured 
photon intensity between 10 and 50\%. As observed in Sect.~\ref{sect:dance}, the 
agreement of simulations with experimental data worsens at higher atomic numbers.
When a global scaling factor is applied in order to match 
the integral intensity of the bremsstrahlung continuum ($> 20$~keV), all models show a
good agreement with the measured shape of the photon energy spectra 
(see Fig.~\ref{fig:ambroseNormalized}). The Livermore and Penelope models 
provide the best agreement to data.
%
\section{Conclusions} \label{sec:conclusions}
The present work shows that all bremsstrahlung physics models provided by \geant\ are in 
reasonable agreement with the experimental data below 3~MeV, both for absolute yields and 
for spectral shapes. Given the data uncertainties and the lack of a wider set of measurements 
from different sources, it is hard to provide any deeper assessment of the discrepancies that 
have been observed. However, a few generic conclusions can be drawn.
Good agreement is obtained when forward photon emission is considered, while the agreement is spoiled 
for back-scattered photons, that are over-estimated both in number and in energy.
As a general trend, all physics models show a better agreement with measurements at 
lower Z values and higher incident electron energies. A good agreement is also observed for 
the shape of the photon energy spectra emitted by electrons at 70~keV. The total radiated 
energy is slightly under-estimated by the \geant\ models, the difference ranging from 
10--20\% in light elements up to 50\% in heavy elements. Such a level of precision can be 
considered as satisfactory in this energy range.\\
The Penelope model shows the best agreement with measurements 
for incident electrons energies in the MeV range, but the choice is much less evident 
at lower energy (below 100 keV). Nevertheless, since in medical physics applications the 
electron kinetic energies are spread over a wide range, from some keV to tens of MeV, 
it should be generally preferable to use a physics model able to give an average better 
response over the range of energies of interest. Consequently, according to our qualitative 
assessment, the Penelope bremsstrahlung model appears to be slightly preferable with respect 
to the other two available in \geant. \\
The agreement between \geant\ simulations and experimental data is found in this work to improve 
as energy increases. However it is still important to provide a direct validation of the 
physics models in the energy range of medical applications (3--10~MeV), where experimental data 
are scarce. Since direct measurements of the spectral distributions are difficult, some authors 
performed in-phantom measurements of the bremsstrahlung dose by using an electron 
``raw beam''~\cite{faddegon2009,sawkey2010}.
Additional data in the energy range of 3--10~MeV will eventually give more information to 
potentially improve the models and consequently the accuracy in the Monte Carlo medical physics 
applications.



\bibliographystyle{elsarticle-num} 
\bibliography{biblio}






\end{document}